# Decentralized Clinical Trials in the Era of Real-World Evidence: A Statistical Perspective

Jie Chen[1] , Junrui Di[2], Nadia Daizadeh[3] , Ying Lu[4], Hongwei Wang[5], Yuan-Li Shen[6] , Jennifer Kirk[6], Frank W. Rockhold[7] , Herbert Pang[8] , Jing Zhao[9], Weili He[5], Andrew Potter[6], and Hana Lee[6]*

[1]Data Science, ECR Global, Shanghai, China

[2]Global Product Development, Pfizer Inc, Cambridge, MA, USA

[3]Advanced Quantitative Sciences, Novartis Pharmaceuticals Corporation, East Hanover, NJ, USA

[4]Biomedical Data Science, Stanford University, Stanford, USA

[5]Data and Statistical Sciences, AbbVie, North Chicago, Illinois, USA,

[6]Food and Drug Administration, Silver Spring, Maryland, USA

[7]Department of Biostatistics and Bioinformatics, Duke University Medical Center and Duke Clinical Research Institute, Durham, North Carolina, USA

[8]PD Data Sciences, Genentech, South San Francisco, California, USA

[9]Biostatistics and Research Decision Sciences, Merck & Co., Inc., North Wales, Pennsylvania, USA

## Funding Information

There is no funding support to this research.

## Conflict of Interest Statement

The authors report no conflict of interests. The FDA had no role in data collection, management, or analysis. The views expressed are those of the authors and not necessarily those of the US FDA.

*Correspondence to: Dr. Hana Lee, CDER, FDA, Silver Spring, MD 20903. Email: hana.lee@fda.hhs.gov.

# Decentralized Clinical Trials in the Era of Real-World Evidence: A Statistical Perspective

### Abstract

There has been a growing trend that activities relating to clinical trials take place at locations other than traditional trial sites (hence decentralized clinical trials or DCTs), some of which are at settings of real-world clinical practice. Although there are numerous benefits of DCTs, this also brings some implications on a number of issues relating to the design, conduct, and analysis of DCTs. The Real-World Evidence Scientific Working Group of the American Statistical Association Biopharmaceutical Section has been reviewing the field of DCTs and provides in this paper considerations for decentralized trials from a statistical perspective. This paper first discusses selected critical decentralized elements that may have statistical implications on the trial and then summarizes regulatory guidance, framework, and initiatives on DCTs. More discussions are presented by focusing on the design (including construction of estimand), implementation, statistical analysis plan (including missing data handling), and reporting of safety events. Some additional considerations (e.g., ethical considerations, technology infrastructure, study oversight, data security and privacy, and regulatory compliance) are also briefly discussed. This paper is intended to provide statistical considerations for decentralized trials of medical products to support regulatory decision-making.

**Key words**: digital healthcare technology, estimand, remote data acquisition, statistical analysis plan.

## 1    Introduction

The U.S. Food and Drug Administration (FDA) defines a decentralized clinical trial (DCT) as a clinical trial where "some or all of the trial-related activities occur at locations other than traditional clinical trial sites".[1] Unlike traditional clinical trials (TCTs) that are usually supported by a specific research infrastructure (e.g., TCT sites), DCTs are intended and designed to reach participants beyond the TCT sites, with potential to improve participation in clinical trials and allow for continuous data capture in real-world settings.[1,2] With increasing use of digital healthcare technologies (DHTs) in clinical investigations, DCTs can facilitate participation by more diverse patient populations in various community settings where healthcare is delivered and can generate information that is more representative of the real world and may help patients and healthcare providers make more timely informed treatment decisions. [3] The other benefits of DCTs may include, e.g., faster accrual and improved retention of participants, reduced burden to both participants and sponsors, increased representativeness of the target population, and generating real-world effectiveness of medical products. [2,4,5] Engaging broader participation of healthcare providers into trials also has the potential to accelerate implementation of successful treatments from publications and approvals to the actual use in patient care.

In general, DCTs can be characterized by reduced operational reliance on specialized research facilities and intermediaries for *trial conduct* (e.g., administration of an investigational product or IP at locations that are convenient for participants and disease assessment by local healthcare providers or telemedicine) and *data collection* (e.g., remote data collection via DHTs in a clinical investigation). Depending on the degree of decentralization, a DCT can be fully decentralized where all activities take place at locations other then TCT sites (e.g., at the homes of trial participants or in local healthcare facilities that are convenient for trial participants) or a hybrid DCT where some trial activities require participant's in-person visits to TCT sites and other activities are conducted at locations that are convenient to participants. [1]

DCTs can accommodate various trial designs such as TCTs and pragmatic clinical trials (PCTs),[6–8] with decentralization referring to attributes related to the methods and procedures governing the conduct of clinical investigations and not study design features. [2] For example, many TCTs routinely apply hybrid locality approaches for data generation and collection (e.g., specimen samples shipped to off-site locations for testing and semi-virtual data collection methods used to obtain protocol-specified information through email, facsimile, telephone, or other DHTs), [2,9] and PCTs, often involving some secondary use of data collected from subjects in routine care, [10] can also leverage DHTs to conduct the trial at off-traditional sites and collect trial data during clinical practice. [11] Therefore, in the continuum of clinical investigations from TCTs conducted in highly controlled clinical settings to PCTs embedded in routine clinical practice, decentralization can be used as a method for remote conduct and monitoring of trial-related activities (including remote data collection). The relationship among TCTs, PCTs, and DCTs can be illustrated in Figure **1.**

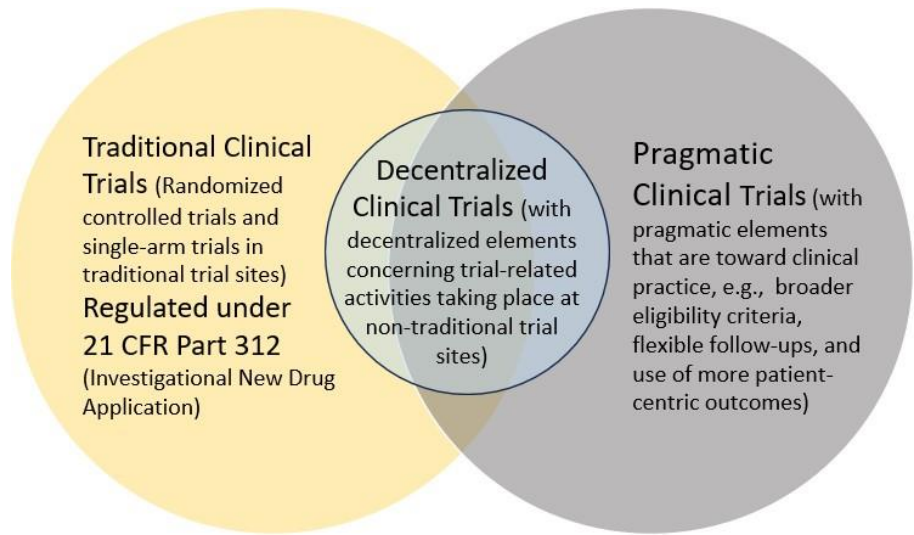

Figure 1: Relations among traditional clinical trials regulated under 21 CFR Part 312 (Investigational New Drug Application), decentralized clinical trials, and pragmatic clinical trials.

Although DCT designs can be applied to all trials of any disease areas, they are particularly suitable to trials for chronic diseases, rare diseases, immobile participants, self-administered IPs, and lower-risk-profile products. [1,4,12,13] While clinical trials can be decentralized at various levels in different scopes, there are some common design features that are worth of further discussion from statistical perspectives to ensure that the DCT design (including data acquisition) and analysis are appropriately aligned with the study objectives and corresponding estimands. Toward this goal, the Real-World Evidence (RWE) Scientific Working Group (SWG) of the American Statistical Association (ASA) Biopharmaceutical Section (BIOP), under the auspice of the Public Private Partnership (PPP) of the Center for Drug Evaluation and Research (CDER) in the FDA, has spent significant effort to assess the landscape of DCTs in medical product development and regulatory decision-making. With this in mind, this paper aims to (1) describe selected critical decentralized elements of clinical trials that are important from a statistical perspective, (2) review relevant regulatory guidance documents pertaining to DCTs, and (3) discuss statistical considerations for the design, conduct, and analysis of DCTs.

This research first describes selected key Elements of DCTs that are statistically relevant, followed by worldwide Regulatory Guidance and Framework on DCTs. Then Statistical

Challenges and Considerations for the design, conduct, analysis, and reporting of DCTs are presented. The paper concludes with a Discussion and Concluding Remarks.

## 2 Elements of DCTs

There are many decentralized activities that can be incorporated into a clinical trial. However, not all of these elements have a direct impact on the statistical aspects of the trial, e.g., management of source documents at decentralized sites may have a minimum impact on the design, analysis, and result interpretation of the trial. This section is intended to discuss selected decentralized elements that may have a potential impact on the trial from a statistical perspective.

### 2.1 DHTs

DHTs play crucial roles in enabling and facilitating decentralized trial activities such as telemedicine and remote data acquisition and monitoring. DHTs include technologies (e.g., wearable, implementable, ingestible, and environmental sensors) and software applications (e.g., connection and computing apps) on portable/mobile devices (e.g., smart-phones, smart-watches).[14,15] Use of DHTs for continuous data acquisition in free-living environments allows the capture of pre-specified measurements (e.g., activity level) as well as a new category of objective measurements that may illuminate the nature of disease progression and derives clinically meaningful endpoints that were previously impossible.

*Choice of fit-for-purpose DHTs.* There is a wide spectrum of DHTs for potential use in DCTs. The choice of a DHT should be fit-for-purpose, e.g., the DHT needs to be suitably qualified and validated for its intended use (e.g., evaluation of endpoints based on data captured by DHTs) to provide reliable data that can produce interpretable results via appropriate statistical analyses when submitted to the regulatory agency for decision-making. The FDA guidance on DHTs[14] provides considerations on the selection, verification/validation, evaluation of endpoints involving data captured by DHTs, statistical analyses, and risk management and prevention when using DHTs in clinical trials. The Digital Medicine Society (DiME) proposed the V3 framework (verification, analytical validation, and clinical validation) as an industry gold standard to evaluate if a DHT and the corresponding digital endpoint are fit-for-purpose in clinical trials for any specific indication.[16] Note that ensuring all participants have access DHTs or other technologies including "bring your own device (BYOD)" is critical for increasing participation, compliance, and diversity of participants, which may lead to a high variability of collected data. [1,17]

*Artificial intelligence.* The FDA guidance on DCTs explicitly mentions that software installed in DHT devices may include those enabled by artificial intelligence (AI).[1] In addressing challenges related to the use of DHTs in regulatory decision-making, the FDA framework for the use of DHTs in product development points out that DHTs may incorporate validated and fit-for-purpose AI algorithms and models (including machine learning) into drug development such as participant recruitment, site selection, trial data collection and analysis, and safety monitoring. [15] Specifically, AI-enabled DHTs may help a DCT, e.g., (1) identify and enroll eligible participants by finding matching candidates in patient databases (e.g., patient registries) (see also Participant screening, recruitment, and retention), (2) analyze a set of completed clinical trials and related databases and assess how to adjust eligibility criteria for broader participation, (3) send (or not send) a customized message or a deadline reminder to participants for taking trial medications and/or completing electronic clinical outcome assessment (eCOA) (e.g., taking and sending photos and videos of diseased area for decentralized assessment), (4) continuously collect temporal data in a typical living environment that may not be captured in site-based trials, and (5) predict the success probability based on drug molecule, target disease, and participant eligibility criteria. See Liu et al.[18], Thomas and Kidziński[19], Chen et al.[20], FDA[21], Harmon et al.[22], Hutson[23], NMPA[24], and references therein for more discussions on the use of AI and DHTs in DCTs.

*Estimands.* The FDA guidance on DHTs[14] points out that for late phase trials that use DHTs for data acquisition, the study protocol and statistical analysis plan (SAP) should follow the estimand framework of ICH[25] and discuss the potential impact of possible events associated with the use of DHTs (e.g., malfunction of DHTs after treatment initiation) for data acquisition and interpretation of endpoints. The SAP should specify how these events will be handled when estimating the treatment effects based on the data of endpoints collected via the DHTs. See Estimands for more discussions on considerations of defining estimands in DCTs.

## 2.2 Participant screening, recruitment, and retention

Leveraging DHTs to remotely identify, screen, enroll, and retain participants is common practice in DCTs. For example, a DCT may use DHTs for

- *digital advertisement and trial promotion* by reaching out potential participants, especially those who otherwise do not have access to the trial information (e.g., those living in remote areas), via a wide range of tools such as personal devices (e.g., smartphones, smart watches) with appropriate apps installed (e.g., ObvioGo, TrialOS); [26]

- *identification of potentially eligible participants* by using some screening tools such as online questionnaires or reviewing individual's EHRs/EMRs to ensure that potential participants meet trial eligibility criteria (those who are potentially eligible for a trial will go through further verification before signing the informed consent); [27,28]

- *remote informed consent* of confirmed eligible participants by using web-based tools for signature. It is important to ensure that participants are fully aware of the details in the informed consent including the risks, benefits, and possible alternatives by participating in the trial. [28,29] For example, the REMOTE trial uses a multiple-choice test to confirm whether participants fully comprehend the informed consent before signing it; [30]

- *participant's engagement and retention* by (1) remote instructions for use of IPs and laboratory test kits that can be sent to participants' location for off-site collection and testing of biosamples (the trial protocol may need to define standardized procedures for handling, testing, and reporting of off-site laboratory tests), (2) telemedicine visits for study follow-ups, assessment, and consultations, (3) continuous support through multiple channels (e.g., text message, emails) to keep participants engaged and informed about trial progress; [28,31]

- *data collection* through implementing secure and user-friendly electronic data capture (EDC) systems that allow participants to enter data (including electronic patient-reported outcomes or ePROs) remotely from home-based terminals or portable devices (e.g., laptops, smartphones or smart watches) and data monitoring through online monitoring system for data quality and compliance. [32,33]

Each of the above items may potentially affects the broadness and compliance of participants and the quality and reliability of data captured during the trial. For example, use of appropriate devices and apps will certainly reach out broader populations and enroll more diversified groups of participants, especially those who live in remote areas with limited access to TCTs, while helping improve the quality of trial conduct and collected data. [20] Of note, recent technology advancement provides new tools to minimize fraud and duplicated enrollment (e.g., for compensation purpose) by verifying participant's credential and eligibility. [34]

## 2.3 Dispensing medication

In addition to the considerations regarding packaging and shipment of IPs as described in the FDA guidance on DCTs, [1] the trial sponsor needs to ensure during site selection and due diligence that the partnering local or mobile HCPs are able to administer the medications which can be delivered to participant's verified address using courier services. The feasibility of shipping medications to trial participants has already been demonstrated during the COVID-19 pandemic. [30,35] Participants will be asked to confirm receipt of trial medication and the condition of contents and have them clearly recorded in the case report form (CRF). If the IP is required to be administered by a HCP, then shipping directly to the participants may be inappropriate. For self-administered IP, sufficient, easy-to-understand, and step-by-step instructions should be provided to the participants to avoid medication error. [1,36] Of note, it is strongly recommended that some customized reminding messages be sent to participants regarding when and how the self-administered medication should be taken; see also DHTs.

## 2.4 Outcome/endpoint assessment and data acquisition

Using DHTs (e.g., telemedicine) and/or local HCPs for remote outcome/endpoint assessment and data acquisition is one of the most important features in DCTs. The FDA guidance on DCTs[1] points out that the data collected in a DCT may be more variable and diversified and hence less precise than those obtained in a site-based TCT. This may present challenges for designing a decentralized non-inferiority trial when the non-inferiority margin is derived from data collected from site-based studies. Depending on decentralized elements used for outcome/endpoint measurements (e.g., measurements at home by participants or HCPs, via a remote electronic or mobile device, tests performed at local laboratory facilities), data collected using DHTs may lead to data quality concerns such as missing data (e.g., due to compliance with DHTs, especially for long-duration DCTs), biases (e.g., self-preferred PROs), reliability (e.g., unreliable laboratory test results caused by incorrect collection of biosamples), traceability/auditability (e.g., no written forms of source data), and secure storage and transmission;[37] see also Section Data management and monitoring. Additional challenges may also include (1) the potential impact of software updates in DHT platforms used for data acquisition over time (e.g., comparability of data collected between two different versions of software), (2) possibly skewed data collected if a DCT enrolls a high proportion of tech-savvy, younger participants, (3) infeasibility of validating virtually collected data, and (4) difficulty in interpretation of a large volume of data, especially from different sources.[4]

Given the above challenges in outcome/endpoint assessment and data acquisition, regulatory guidance documents (see Section Regulatory Guidance and Framework on DCTs) state that the study protocol should specify the rationale for decentralized strategies that make the trial more convenient and accessible to trial participants and measures to ensure data quality and reliability (e.g., use of local HCPs to reducing missing data due to DHT defects or poor technology adherence). Note that methods used for outcome/endpoint assessment and data acquisition should be chosen not only to be able to answer the clinical question of the trial, but also to reflect the interest and/or preference of the trial participants. With technology advancements, a broad range of clinical outcomes/endpoint can be collected using home-based assessments. In the absence of such technology, community-based assessments such as local HCPs and local laboratories should be leveraged (Figure 2). Note that most clinical outcomes may need to be assessed by physicians or, sometimes, specialists.

## 2.5   Safety monitoring

With AI-powered DHTs, participants can be monitored in real-time during the DCT for timely flagging and alerting safety issues. For example, wearable ECG patches can detect a wide range of cardiac events including previously undiagnosed atrial fibrillation.[38] The FDA guidance on DCTs[1] specifies that the trial protocol should include a safety monitoring plan describing how the participants are expected to respond to and report adverse events (AEs) and seek for medical assistance if needed. Some key considerations for safety monitoring in DCTs may include: (1) efficient use of DHTs for remote monitoring (e.g., collection of data on vital signs, symptoms, medication adherence), (2) integration of various sources of data with advanced analytics for prompt detection of possible safety signals, (3) use of centralized safety oversight to review safety data and make recommendations on a regular basis, and (4) adaptive safety strategies based on emerging safety issues. In addition, careful considerations should be given to (1) procedures for communication, documentation, and implementation of self-monitoring and self-reported AEs, (2) possibilities of participant's inability or difficulty to report such events due to, e.g., cognitive impairment or lack of connectivity, (3) risks of erroneous reporting due to, e.g., malfunction of the DHTs, and (4) subsequent steps for participants who report serious safety concerns (possibly serious adverse events or SAEs) to take advanced medical examination that otherwise cannot be done by local HCPs. An example of decentralized safety monitoring, augmented by traditional site-based data capture, is the REACT-AF study [39] in which hospitalizations as one of PROs are monitored and verified in nearly real time from participants' hospitals with connection to their EHR and transfer of hospital discharge summaries to study team. Of note, the sponsor must cease the remote administration of an IP if it is associated with significant risks or serious AEs.

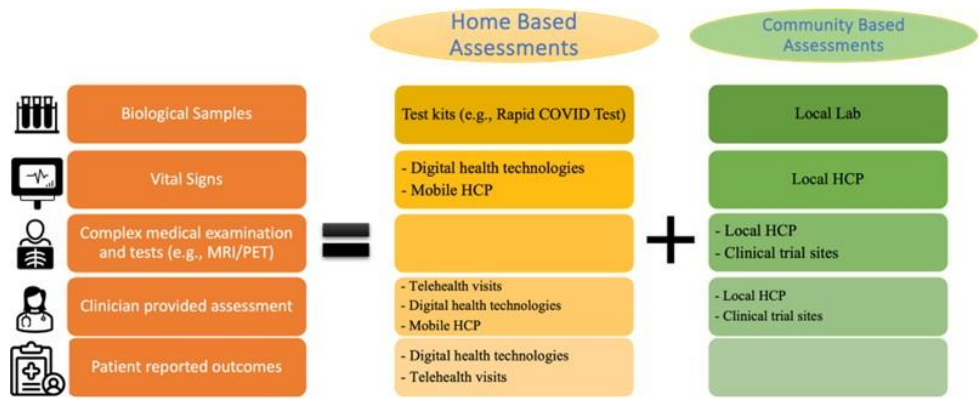

Figure 2: Fit-for-purpose data collection and clinical assessments.

## 2.6 Data management and monitoring

The volume and complexity of data collected from a DCT are quite different from those obtained from a TCT. However, the regulatory standard for data quality and reliability remains unchanged for data submitted in support of regulatory decision-making. The regulatory guidance documents on DCTs such as the FDA guidance on DCTs[1] recommend that a data management plan (DMP) include at least (1) data origin and data flow from all sources, (2) methods used for remote data acquisition from all sources (e.g., DHTs, local HCPs), and (3) a list of vendors involving in data collection, handling, and management. In addition, it would be helpful to consider in more details the following aspects in the DMP of a DCT:

- *DHTs and their management during the trial.* Special attention should be given to the following five categories of more data-related DHTs: (1) disease diagnostics (e.g., measuring disease status, progression, response, or recurrence), (2) therapeutics (e.g., by generating or delivering a medical intervention), (3) monitoring and tracking (e.g., monitoring specific health conditions or tracking participant's performance), (4) care support (e.g., self-management of a medical condition through education, recommendations, or reminders), and (5) health system (e.g., providing HCPs with a tool to manage their patients). These categories of DHTs will generate different types of trial data, some of which could be indicated in the product insert and hence are of great importance for the subsequent analysis and regulatory decision-making. The management of DHTs (including V3, software updates, malfunction, replacement, etc.) should be detailed in the DMP.

- *Sources of data.* Considering multiple sources of data to be generated, the DMP should describe (1) specific data points to be generated by a DHT, participants, or local HCPs, (2) the quality control system to ensure consistency and reliability of the data generated by different tools (especially by BYOD), and (3) processes for data storage, transfer, and anonymity.

- *Data review.* A coordinated plan should be defined and implemented to establish the scope and process of data review by individual critical functions such as (1) safety specialists for subject safety, (2) clinicians for monitoring efficacy and/or safety end-points, (3) data managers for data consistency, logic errors, cleanness, and (4) data analysts for outliers, trend, patterns, missing data, etc. It is critically important to focus on standardization and consistency due to multiple sources of data to be collected.

- *Real-time analytics.* Using advanced analytic methods built into the DHT system can greatly improve the efficiency of data management. The DMP may describe how analytic tools can be used in a real time, e.g., (1) to generate more sophisticated visualization and tabulation for monitoring participant's safety, data quality, and trial conduct, (2) to perform pre-planned analyses for diagnosis, pattern detection, and prediction,[40] and (3) to combine multiple sources of data for integrated summaries of efficacy and safety.

In summary, the DMP of a DCT should discuss in more details the acquisition, handling (e.g., transmission, security, and privacy), and management of data that are generally in a large volume, from multiple sources, at a great complexity and variability, and with a variety of challenges in data generation.

## 3  Regulatory Guidance and Framework on DCTs

It is important to understand the current positions, recommendations, and/or other considerations by regulatory agencies to ensure that DCTs are designed and conducted under relevant regulatory framework. The following briefly describes the guidance and recommendations on DCTs by regulatory agencies worldwide.

Realizing continued evolvement with novel trial designs and technological advancement, the International Council for Harmonisation of Technical Requirements for Pharmaceuticals for Human Use (ICH) revised its technical guidance on Good Clinical Practice (GCP) to accommodate trial activities to be taken place in decentralized settings.[41] The ICH E6(R3) guideline states that the scientific integrity of the trial and the reliability of trial results depend on the trial design, which should include, among others, a description of the type and design of trials to be conducted (e.g., trials with decentralized elements). The Annex 2 of the ICH GCP E6(R3) states that the GCP principles are applicable across a variety of trial designs (including DCTs, PCTs) and data sources (including real-world data).[11]

Meanwhile, individual regulatory agencies worldwide, such as the US FDA, the European Medicines Agency (EMA), the National Medical Products Administration (NMPA) of China, Swedish Medical Products Agency (SMPA), Denmark Medical Agency (DKMA), and Swissmedics), have issued their own guidance documents and/or recommendations for the design and conduct of DCTs. Although the scope, contents, and recommendations are slightly different in the guidance by different agencies, they all share the same principles and general considerations, which can be summarized as follows:

- *Rationale of DCTs*—The scientific, operational, and regulatory basis for a trial to be designed and conducted as a decentralized trial, e.g., some unpreventable causes such as COVID-19 pandemic, enhancing trial participation by subjects with limited mobility, and improving participant engagement, recruitment, and retention.

- *General considerations for designing and conducting a DCT*—The scope, planning, and implementation of a DCT such as (1) the degree of decentralization (e.g., fully decentralized or hybrid decentralized), (2) a specific plan to implement the decentralized elements (e.g., use of local HCPs or DHTs), (3) specific issues related to the feasibility, design, implementation, and data collection and analysis of the trial, (4) description of appropriate training, oversight, and pre-defined risk management and mitigation plan when implementing a DCT, and (5) an *a priori* discussion/communication with and agreement by relevant regulatory agency.

- *Use of DHTs*—A description on (1) the details of DHTs to be used in the trial (including qualification and usability of the DHTs), (2) utility of the DHTs (e.g., compliance monitoring, data collection and transmission), (3) training on the use of DHTs, and (4) risk management and mitigation related to the use of DHTs in the trial.

- *Decentralized elements*—Most remote conduct of trial-related activities such as (1) electronic informed consent from participants at their locations, (2) remote clinical visits (telemedicine), (3) use of local laboratory facilities, (4) administration of IP at locations convenient for participants, (5) remote site monitoring, (6) safety monitoring, and (7) remote source data verification.

A comparison of guidance and/or recommendations on the general considerations on decentralized elements by different regulatory agencies worldwide is presented in Table 1 of the Appendix, which may help design multi-regional decentralized trials for submissions to multiple regulatory agencies.

## 4 Statistical Challenges and Considerations

Decentralization brings unique challenges and require careful considerations in the design, conduct, and analysis of a DCT to ensure validity, robustness, and reliability of study results. This section discusses selected challenges and corresponding strategies or considerations to address them from a statistical perspective.

### 4.1 Estimands

An estimand connect the study objective with the target of inference and drives all subsequent steps (including design, conduct, analysis, and result interpretation) of a trial. FDA guidance on DHTs[14] points out that late phase studies should use the estimand framework of ICH [25] to precisely define the estimand of the study—the treatment effect to be estimated. Although there is a similarity in estimand construction between DCTs and TCTs, some additional considerations are worth of further discussion to address the special features and challenges associated with decentralized designs, remote data acquisition, and possibly diverse participant populations.

(a) *Population*. Participant populations in DCT may differ from those in TCT on the following aspects: (1) geographically more dispersed (e.g., more participants from remote regions), (2) more inclusive and accessible to broader participants (e.g., greater participation of physically immobile participants such as those with more severe conditions or with transportation limitations), (3) attracting those who are tech-savvy or preferring remote participation, and (4) increased retention and adherence for those who prefer flexibility and convenience.

(b) *Treatment*. Some special considerations for treatment and its delivery options in DCTs may include: (1) remote delivery of treatment (including self- and HCP-administered treatment), reliability, and compliance, (2) behavioral factors affecting treatment adherence, (3) monitoring and recording compliance to treatment strategies, (4) dynamic adjustment of treatment due to, e.g., self-perception, immediate reaction to possible AEs, or self preference, and (5) any other reasons causing medication errors (e.g., mishandling of IP shipment).

(c) *Endpoints*. Endpoints should be appropriately selected to reflect treatment effect (or its pathway) on the health condition and assessment of the endpoints should also be as precise and accurate as possible. Because of remote data acquisition in DCTs, some additional considerations on endpoint selection and assessment should be taken: (1) incorporation of participant's interest and preference into endpoint selection, (2) using validated and reliable tools (e.g., wearable devices) for endpoint ascertainment and validation, (3) standardized procedures for telemedicine and virtual assessment of outcomes to ensure consistency, (4) considering composite endpoints to combine multiple relevant component endpoints to measure treatment effect, (5) possibly high variability of endpoint measurement if BYOD is used, and (6) relationship with previously established endpoints that have been used to support regulatory decision for similar indications, (7) feasibility to the intended population (e.g., elderly participants who are unable to use a particular DHT to report PRO), (8) choice of appropriate methodologies for clinical validation of the digital endpoints (e.g., Rego et al.[42]), and (9) acceptability by regulatory agencies. Of note, hybrid DCTs may produce a set of endpoints, some of which are captured remotely and some others are on-site.

(d) *Intercurrent events* (ICEs). Besides ICEs that are associated with intolerability (e.g., serious AEs) and lack of efficacy of assigned treatment and terminal events (e.g., death, amputation), participant's personal behavioral factors (e.g., personal preference, friend's recommendation) that may cause discontinuation of the assigned treatment should also be accounted.[43] Note that this category of ICEs is essential for patients, healthcare providers, and regulators to make informed decision about product effectiveness when complying with the treatment regimen. Of note, malfunction of DHTs that are used to measure endpoints for estimating treatment effects may also be considered as an ICE if it causes interruption of assigned treatment continuation. However, it is important to distinguish the "informed presence" and "informed non-presence" of missing values of endpoints so that appropriate strategies can be applied to address them. [44]

(e) *Population-level summary.* Conceptually, the difference in population-level summary between a DCT and TCT should be minimal as treatment efficacy and effectiveness may be expressed similarly, but under different application settings.

Given the above considerations, appropriate specification of estimand attributes is important to ensure the clinical questions can answered precisely and accurately. Izem et al. [45] suggest that (1) sponsors need to consider whether the DCT is targeting a novel estimand, rather than an estimand that was previously targeted by clinical trials with on-site components, (2) it is crucial to decide whether an estimator from a DCT can provide an improved estimate with respect to the target estimand in terms of validity, fitness-for-purpose, potential bias, and precision, and (3) it is important to consider the extent of alignment between the design and the question on external validity, or generalizability of findings. Such considerations tailored to DCTs will enable the ramification of decentralization on the estimand elements—population, treatment, variable, and ICEs.

## 4.2 Trial design

A good practice in clinical studies is to incorporate anticipated challenges into the study design so that the collected data can be used to answer the clinical question of interest.

*Participant heterogeneity and outcome/endpoint variability.* Although the target population may remain unchanged (except for PCTs which may have less strict eligibility criteria), participants in DCTs may well be diversified compared with TCTs, as discussed in the population attribute of estimands in Estimands. This will lead to a higher degree of participant heterogeneity with respect to demographics, phenotypes, and genotypes, lead to possibly more variability of outcome/endpoint measures, which could be particularly true for DCTs with pragmatic elements. [5,46,47] A possible strategy to address participant heterogeneity and outcome/endpoint variability is to incorporate them in the trial design (e.g., stratified randomization, appropriate sample size estimation) and data analysis (e.g., stratified analysis, pre-defined covariate adjustment). [48]

*Sample size determination.* The principles of sample size determination for DCTs are similar to those for TCTs. However, some unique considerations may be needed due to decentralized nature. First, using effect size estimates derived from prior site-based TCTs may lead to an under-powered DCT if the latter incorporates pragmatic elements that may enroll more heterogeneous participants and use endpoints measured in real-world setting that are often more variables as discussed above. Second, participants who choose to be decentralized may be clustered within study sites, region, or other study units, leading to intra-cluster correlation (ICC) when cluster randomization is used. Ignoring this ICC may result in an under-powered study.[49] Third, although decentralization may help improve participation, some other factors such as literacy in DHTs, access to internet, and continuous monitoring/collection of data may affect attrition rates differently for participants between comparison groups.

*Randomization.* Willingness to remote (or on-site) participation may differ among different subpopulations, e.g., younger participants or those living in remote areas may prefer decentralized participation as compared with older participants or those who reside in urban areas. Adaptive or stratified randomization may be considered to account for possible imbalance between treatment groups within some strata.

*Blinding.* Decentralization may increase the risk of unblinding treatment allocation in blinded DCTs. For example, the following factors may possibly cause inadvertently unblinding: technology breaches (e.g., malfunction of a DHT), use of BYOD (e.g., compatibility or unnoticed glitches), insufficiently trained local HCPs (e.g., failure to recognize importance of blinding), noncompliance of participants (e.g., individual's expectation, preference), and other logistic reasons (e.g., mishandling of IP shipment to individual participants, error in shipping, labeling, and packaging). The above challenges that may compromise blinding should be addressed in the study protocol at the design stage. Of note, a simpler procedure of blinding is always preferred (and should be specified in the protocol) to ensure that investigators (including HCPs) and participants can fully understand and easily comply with the trial protocol.

*Potential biases associated with decentralization and possible measures to minimize them.* Although DCTs may enroll more representative participants of the target population, the nature of remote conduct may induce several types of biases, especially when pragmatic elements (e.g., treatment delivered by HCPs, flexibility for compliance, follow-up through clinical practice, outcome measured via diverse methods, etc.) are implemented. For exam-ple, (1) *performance bias* may occur if treatments under comparison are delivered differently by HCPs or participants themselves, especially if the IP is new to HCPs and the CP is an SoC, (2) *assessment bias* may occur if outcomes/endpoints are assessed/reported differently by HCPs who are relatively inexperienced with the IP and associated medical conditions, (3) *attrition bias* if participants discontinue the assigned treatments in different rates or patterns between comparison groups with the IP is associated with inconvenient use or any particular intolerable adverse events compared with the CP. To reduce these potential biases, a well-designed training plan can be implemented to provide detailed instructions to HCPs on delivery of study products and assessment of outcome/endpoints and to participants on study compliance and outcome reports (e.g., ePROs).

## 4.3 Statistical analysis plan

The FDA guidance on DHTs [14] recommends that analyses of data collected from DHTs be discussed in the statistical analysis plan (SAP) of a DCT, which should include the endpoint under consideration, the IP under investigation, and the study population in which the product will be used upon approval. In particular, the Guidance points out that the SAP should discuss: (1) the methods used for data collection, (2) inappropriateness for decentralized non-inferiority (NI) trials if the NI margin is derived from non-decentralized studies, (3) pre-specified endpoints and source data from which the endpoints are derived, (4) using estimand framework to precisely describe treatment effect (see Estimands), and (5) events and issues that may affect data collection, data quality, missing data, and subsequent analyses.

In addition to the common components of an analysis plan (e.g. detailed description of estimands, analysis methods, handling protocol deviations), the SAP may also consider further discussion on strategies and methods of handling heterogeneity of endpoints. In general, the variability of endpoints derived from data captured by DHTs in DCTs may increase due to reasons such as (1) more diverse participants (e.g., by eliminating geographical and transportation barriers that often limit participation of under-represented populations) with different access to and knowledge of technology, socioeconomic status, and cultural backgrounds that may impact the reliability and consistency of endpoint measurement, (2) use of local HCPs (e.g., endpoint assessment by individual HCPs) and laboratories (e.g.

biomarker endpoints tested at local laboratories), (3) self-measured disease conditions (e.g., self-taking and -uploading images/videos for endpoint measurement) and self-reported outcomes (e.g., ePROs), (4) use of different devices or platform (e.g., BYOD) that may lead to inconsistency in data acquisition, and (5) some other factors (e.g. accessibility to local healthcare, different regulations and healthcare systems) that may contribute to the increased variability of endpoint data.

The aforementioned multiple sources of data heterogeneity can bring some challenges for statistical analyses that should be addressed in the SAP. In particular, under the estimand framework discussed in Estimands, the analysis plan may need to consider: (1) the impact of heterogeneous endpoint data on the precision of estimated parameters with appropriate sample size and statistical power, (2) using appropriate models to incorporate different sources of measurement variabilities, e.g., covariate adjustment modeling, within- and between-subject variances for longitudinally captured endpoint data, (3) including terms of biases possibly caused by decentralized measurement in hybrid DCTs, e.g., Curtis and Qu[37], and (4) appropriate estimators with confidence intervals and coverage probabilities to ensure reliability, unbiased, and efficiency of estimators. In summary, the SAP should thoroughly discuss possible sources of data heterogeneity and potential biases and strategies to address them to ensure validity and reliability of analysis results.

### 4.4 Missing data

Missing data is a common problem in clinical studies. The general principles for the prevention and treatment of missing data in clinical trials[25,50,51]) are applicable to all types of clinical trials. Although the strategies and methods of handling missing data (including missing data imputation algorithms) should be part of the SAP of a DCT, the importance of missing data deserves a separate discussion in more details, given the special features of missingness in DCTs. First, it is important to understand the mechanisms of missingness in DCTs. Consider the three common missing data mechanisms—missing completely at random (MCAR), missing at random (MAR), and missing not at random (MNAR).[52] MCAR might occur if a device malfunction or data loss occurs during data transfer; MAR might occur (within a subpopulation) if missing data are frequently observed within a subset of participants (say, elderly or female subjects) for whom the device is inconvenient to use;

and MNAR might occur if participants elect not to use the device if they feel very well (or unwell). [53] In addition, missing data may have some hierarchical structures in DCTs, e.g., participants with missing data may be clustered within sites or regions, for which within- and between-site (or region) correlation should be considered to avoid under-estimation of variances and inflation of type I error.

Second, the strategies and methods to address missingness should be based on the reasons of missingness. For example, within-subject imputation may be suitable to use data from complete segments of each day to estimate values in incomplete segments when MAR is assumed (i.e., data from the complete and incomplete segments are in the same distribution). Some commonly used techniques handling missing data include imputation (e.g., mean imputation, multiple imputation with predictive mean matching) and maximum likelihood (ML)-based methods (e.g., ML estimation, full information ML).[50] Di et al.[53] also explore functional data analysis to deal with missing data when they are derived from continuous sensor or wearables and some machine learning (e.g., deep learning) methods to account for complex patterns of missingness and variability in missing data uncertainty. When relying on RWD such as the EHRs in DCTs, see Molenberghs and Kenward[54] for interpolation of longitudinal variables with limited individual level variability, Dalton et al.[55] for imputation based on stratified mean, Goldstein et al.[56] for using informative observations, and Beaulieu-Jones et al.[57], Martʹın-Merino et al.[58], and Cesare and Were[59] for conditional imputations.

Third, the mechanisms of missingness are mostly unknown. Sensitivity analyses are often conducted to evaluate the robustness of study findings to different missing data mechanisms and imputation methods. In particular, evaluating the impact of missing data on estimated treatment effects and associated conclusions should consider a range of plausible scenarios of missing data patterns and a variety of missing mechanisms. [44]

Last (and perhaps the most important), the missing data problem should be carefully and thoroughly considered in the design of a DCT, e.g., specification of ICEs related to DHTs and computational platform, detailed instruction (either separate or part of protocol) on trial conduct and data collection. In particular, the following strategies may help minimize the volume and/or proportion of missing data: (1) clear communication and sufficient training to ensure that all participants (including patients, local HCPs) understand the study protocol and data acquisition procedures, which may help minimize errors and reduce missing data, (2) real-time monitoring to detect missing data for timely intervention (e.g., sending a reminder to participants for follow-ups, fixing malfunctioned devices), (3) remote engagement with participants (e.g., telemedicine, electronic reminders) to encourage compliance with study protocol including data collection. As missing is inevitable, collecting the reason of missingness is one of the important tasks as it will help identify whether the missingness is at random or not at random and help use the correct methods to address missingness.

## 4.5  Study conduct

With careful considerations for the design of a DCT (including specifications of estimands and pre-defined SAP and strategies and methods addressing missing data), the successful implementation of the trial will be critical to ensure participant safety, study integrity, and regulatory compliance. Towards this end, the following aspects may be considered from trial conduct perspective.

*Remote informed consent process*. A streamlined and effective informed consent process needs to be developed to allow for remote consent and ensure that participants understand the study procedures, risks, benefits, and data privacy considerations. Remote informed consent can be conducted effectively via secure electronic platforms, video conferencing, interactive multimedia, etc., for document sharing, review, and (digital) signature collection. It is important to (1) develop concise and easy-to-understand informed consent form, (2) provide opportunities for participants to ask questions before signing informed consent, (3) have confirmation of understanding and remote consent oversight, and (4) maintain detailed records of remote consent process.

*Randomization and blinding process*. With respect to stratified or adaptive randomization as discussed in Trial design, a technique[60] can be implemented to dynamically allocate participants to treatment arms for balancing baseline covariance. However, it is strongly recommended to communicate with relevant regulatory agencies prior to planning and/implementing outcome adaptive randomization in certain disease areas. To maintain participant blinding, regular study integrity checks can be conducted to confirm whether blinding procedures are being followed, which may involve (1) remote surveys or inter-views of participants and/or investigators about their awareness of treatment assignment, (2) monitoring adherence of trial protocol, and (3) investigating any protocol deviations or breaches of blinding.

*Management of DHTs.* DHTs are a critical component in successful implementation of DCTs. Some considerations in managing DHTs in DCTs may include: (1) continuous education and (remote) training (e.g., webinars, training flyers) on the roles and responsibilities of trial staff members about the use of DHTs, (2) quality assurance measures to monitor the performance of DHTs, (3) remote service support to promptly address any technical issues, and (4) flexibility of timely adapting/adjusting for different DHTs if the existing one malfunctions.

*Data monitoring.* The following considerations may help data monitoring in DCTs: (1) data quality checks for inconsistency, logic errors, outlines, and missing data, (2) a risk-based monitoring plan focusing on pre-identified risks to the study, e.g., compliance, critical data elements (treatment, endpoints, and key covariates), (3) monitoring, reporting, assessment, and management of (serious) adverse events, (4) remote monitoring of source documents to assess site performance, (5) centralized, unblinded review on a timely manner of critical data (especially outcome data) by an independent team such as a Data and Safety Monitoring Committee (DSMC). These data monitoring aspects can be highlighted in the study protocol and details of implementation plans can be described in respective documents such as the data management plan and DSMC charter of a DCT.

*Endpoint adjudication.* Many clinical endpoints are assessed remotely in DCTs, which raises the concerns about endpoint consistency and validity. Using independent, possibly centralized, endpoint adjudication committees may help ensure reliability and accuracy of endpoint assessments across sites.

## 4.6   Reporting of safety events

One of the most important features for DCTs is the use of DHTs to facilitate real-time reporting of AEs by participants and remote assessment by investigators. Besides the usually safety reporting requirements as discussed in, FDA [61] additional considerations for safety reporting in DCTs may include: (1) a safety reporting procedure should be in place to detail how AEs (especially SAEs) are collected, documented, reported, and managed,

(2) comprehensive training and educational sessions should be provided to participants and local HCPs on the recognition of AEs, importance and timeline of reporting, and detailed instructions of using remote reporting tools, (3) communication of safety data and findings to participants and trials staff should take into account the diversity of participants with different understanding of safety concerns, and (4) mechanisms of transparently sharing safety data should be established while protecting participant privacy and trial integrity.

## 4.7   Other considerations

*Multi-modal data collections.* DCTs may use multi-modal data collections. Within the same study, it is possible to collect data using multiple wearable devices, eDairies, and clinical outcome assessments simultaneous. This creates another level of complexity in the data standardization and structure. Di et al. [62] provide suggestions on deploying multi-modal sensors in clinical trials which highlights the importance of utilizing the temporal aspects of all modalities and identify the joint effects from multiple modalities. Similar concepts have been emphasized in Zipunnikov et al. [63] as well.

*Regulatory considerations.* In addition to the above discussions and the regulatory guidance documents summarized in Regulatory Guidance and Framework on DCTs, the following additional considerations may help the planning, implementation, and regulatory submission of DCTs: (1) pre-alignment with regulatory agencies on key components of a DCT, e.g., the rationale of using particular decentralized elements, objectives, construction of estimands (precise definitions of each attribute), study design (randomization, blinding, adaptation, etc.), implementation plan to comply with GCP and maintain participant safety and privacy, etc.; (2) timely monitoring collected data to detect any compliance issues (e.g., major protocol deviations), safety concerns, intercurrent events and missing data (particularly associated with use of DHTs), (3) pre-defined statistical analyses to estimate the treatment effect (effectiveness) using strategies (e.g., treatment policy strategy, hypothetic strategy) to appropriately address ICEs that are commonly seen in DCTs or associated with the use of DHTs, and (4) rigorous sensitivity analyses to evaluate the robustness of study findings to the deviation/violation of assumptions based on which the analyses are performed.

# 5    Discussion and Concluding Remarks

Many of the decentralized elements discussed in Elements of DCTs can be found in routine clinical practice and thereby generate RWD that hold great promise for furthering our understanding of medical products and intervention in settings close to the real world. The validity, reliability, and generalizability of study findings from DCTs require sufficient statistical considerations in the design, conduct, and analysis of the trials which have been discussed in this paper. In addition, the success of DCTs also relies on many other factors such as ethical considerations, technology infrastructure, study oversight, data security and privacy, and regulatory compliance. [4,12,64–66] For example, (1) although decentralization offers convenience for participants, there might be some ethical concerns such as inequity of access to technology that should be addressed to reduce disparities among participants; (2) DCTs require reliable and adequate technology infrastructure (e.g., internet connectivity) to support remote data collection and monitoring and traditional EDC systems may need to be modified to handle large volumes of data; (3) remote trial activities require close monitoring and participant engagement to ensure protocol compliance, data quality, and participant safety; (4) measures for secure data transmission and access controls should be established to safeguard sensitive health information; and (5) communication and collaboration with regulatory agencies are critical to address any regulatory concerns specific for DCT designs.

In summary, in contrast with TCTs, DCTs require a number of additional considerations in the design, implementation, analysis, and result interpretation. This paper discusses the major elements of decentralization, relevant regulatory guidance documents or framework, and statistical challenges and considerations from precise construction of estimands, trial design, SAP, missing data, study conduct, reporting of safety events, and some other considerations that may have statistical implications. Some non-statistical related issues are also briefly discussed. We hope this assessment can provide insights into the challenges and statistical strategies and methods to address them toward successful implementation of DCTs.

# Funding Information

There is no funding support to this research.

## Conflict of Interest Statement

## A Appendix

This Appendix provides a tabular comparison of regulatory guidance and/or recommendations on the general considerations and decentralized elements for DCTs by some regulatory agencies worldwide.

Table 1: Comparison of regulatory guidance and/or recommendations for DCTs.

| Agency | General Considerations | Decentralized Elements |
|---|---|---|
| FDA[11] | • Appropriateness for different degrees of (fully or hybrid) decentralized trials<br>• Specific plans to facilitate and implement decentralized elements (e.g., use of local healthcare facilities and remote visits)<br>• Challenges including coordination of trial activities with individuals and facilities<br>• Upfront discussion with FDA on the feasibility, design, implementation, or analysis of a DCT<br>• Variability and precision of data obtained in DCTs, which may not affect the validity of findings of superiority trials, but could affect the validity of findings of non-inferiority trials<br>• Procedures to evaluate and manage adverse events identified remotely<br>• Clearly specified roles and responsibilities for sponsor and investigators | • Electronic informed consent form for participants at their remote locations<br>• Remote trial visits and trial-related activities, e.g., (1) use of local healthcare facilities, local HCPs, and local laboratory facilities, (2) telehealth visits if no in-person interaction is needed, (3) local HCPs to conduct in-person visits and trial-related activities (e.g., performing physical examinations, reading radiographs, obtaining vital signs), and (4) CRFs and other documents to be completed during telehealth visits<br>• Direct distribution of IPs to trial participants at their locations<br>• DHTs for remote data acquisition, management, and transmission<br>• Risk-based monitoring and safety monitoring |
| EMA[36] | • Respect to the rights, safety, dignity, and well-being of trial participants<br>• Adherence to EU and national applicable laws, regulations, established standards and guidance<br>• Engagement with participants, patients (patient organizations), healthcare professionals when designing, developing, implementing a DCT | • Remote informed consent (e.g., use of digital information leaflets, electronic signature)<br>• IPs delivery (from, e.g., pharmacy or a depot) to trial participants and administration at home<br>• A summary of decentralized elements to be provided in the cover letter of trial application |

---

[1]The U.S. Food and Drug Administration

| Agency | General Considerations | Decentralized Elements |
|---|---|---|
| EMA[2 36] | • Description of how decentralized elements to generate reliable and robust data<br>• A contingency plan to minimize the impact of any risk<br>• Procedures in place for reporting and management of adverse events<br>• Description on roles and responsibilities for trial oversight | • Trial-related procedures performed at participant home, e.g., collection, handling, and storage of biological samples<br>• Data collection and management, e.g., direct data capture by trial participants, their caregivers or service providers, electronic systems (e.g., eCRFs, ePROs, wearables, etc.)<br>• Decentralized processes and tools for remote access to and monitoring of trial sites |
| NMPA[3 67–69] | • Principles for patient-centric drug development<br>• Needs-based clinical trial design to include decentralized elements<br>• Improving the experience and reducing the burden of participants<br>• Early communication with regulatory agency to ensure (1) participant opinions to be appropriately adopted and (2) rationality of key outcome assessment<br>• Respect to the rights and protection of benefits of participants<br>• Ensuring data quality and personal information protection | • Online platforms for participant screening and recruitment<br>• IP delivery, storage, handling, and administration<br>• Local study team including remote trial coordinator, remote investigator (for eConsent, COA, safety monitoring, drug administration, etc.), and local HCPs (for physical examination, vital signs, biosample collection, etc.)<br>• Patient-centric activities, e.g., image upload, wearables, ePRO, ePayment, etc.<br>• Remote monitoring and reporting of trial conduct and safety data |

---

[2]The European Medicines Agency
[3]The National Medical Products Administration of China

| Agency | General Considerations | Decentralized Elements |
|---|---|---|
| PMDA[4][70] | • Ensuring protection for safety of trial participants<br>• Ensuring reliability and quality of collected data<br>• Obligation of principal investigator and responsibilities of medical institutions<br>• Training to participants for appropriate use of IPs and compliance<br>• Application of IRB for trial monitoring | • Electronic informed consent<br>• Shipments of IPs directly to participant's home<br>• Self administration of IPs and blood collection for testing at satellite medical institutions<br>• Telemedicine for remote visits and source document verification<br>• Remote GCP inspection (procedure, evidence material, web conference, etc.) |
| CADTH[5][71] | • Encourage decentralization of clinical trials due to dispersed population<br>• Some regulation changes to allow for trial-related activities to be conducted at participant locations | • Documented (instead of "written") informed consent<br>• A witness to attest that informed consent was given<br>• Trial-related activities at participant locations for recruitment, informed consent, monitoring, and virtual visits |
| SMPA[6][72] | • Careful and study-specific risk-benefit assessment<br>• Considerations for DCTs include the type, design, and population of study, characteristics of the IPs, and indication<br>• Appropriate computerized system to handle DCTs | • Remote informed consent process with electronic signature<br>• Remote visits (considering what data are collected and how the results are to be used)<br>• Distribution of PI |

[4]Pharmaceuticals and Medical Devices Agency of Japan
[5]The Canadian Agency for Drugs and technologies in Health
[6]Swedish Medical Products Agency

| Agency | General Considerations | Decentralized Elements |
|---|---|---|
| DKMA[7][73] | • The impact of limited in-person interaction on data quality and trial integrity<br>• Justification and implementation of decentralized elements<br>• Involvement of participants and investigators<br>• Use of new technologies<br>• Choice and validation of endpoints<br>• Plans for remote monitoring of trial participants for compliance and safety<br>• Adverse events reporting<br>• Application requirements by the DKMA | • Subject screening and enrollment through, e.g., social media and established databases<br>• Electronic informed consent via the use of digital systems<br>• Delivery of IPs to participants (storage and transportation) and self administration at home (training and communication)<br>• Remote monitoring of trial participants safety<br>• Digital platforms for data collection, registration and reporting of adverse events<br>• Remote monitoring including remote access to source data |
| Swissmedic[8][74] | • Optimal medical care, the rights and safety of participants<br>• Safe dispensing, ingestion / administration, and returning of IPs<br>• Credible and reliable data recording<br>• Data protection with highest security standards | • Recruitment of participants via digital channels<br>• Trial-related activities performed at non-traditional trial sites<br>• Dispensing and administration of IPs outside of traditional trial sites<br>• Remote source data verification |

---

[7]Danish Medical Agency
[8]Danish Medical Agency

# References

1. FDA. Decentralized Clinical Trials for Drugs, Biological Products, and Devices. Food and Drug Administration, U.S. Department of Health and Human Services, 2023.

2. Khozin, S and Coravos, A. Decentralized trials in the age of real-world evidence and inclusivity in clinical investigations. *Clinical Pharmacology & Therapeutics*, 106(1): 25–7, 2019.

3. Gottlieb, S. Breaking Down Barriers Between Clinical Trials and Clinical Care: Incorporating Real World Evidence into Regulatory Decision Making. Speech at the Bipartisan Policy Center conference, 2019. Accessed: February 20, 2024.

4. de Jong, AJ, van Rijssel, TI, Zuidgeest, MG, van Thiel, GJ, Askin, S, Fons-Mart´ınez, J, De Smedt, T, de Boer, A, Santa-Ana-Tellez, Y, and Gardarsdottir, H. Opportunities and challenges for decentralized clinical trials: european regulators' perspective. *Clinical Pharmacology & Therapeutics*, 112(2):344–352, 2022.

5. Goodson, N, Wicks, P, Morgan, J, Hashem, L, Callinan, S, and Reites, J. Opportunities and counterintuitive challenges for decentralized clinical trials to broaden participant inclusion. *NPJ Digital Medicine*, 5(1):1–6, 2022.

6. Schwartz, D and Lellouch, J. Explanatory and pragmatic attitudes in therapeutical trials. *Journal of Chronic Diseases*, 20(8):637–648, 1967.

7. Loudon, K, Treweek, S, Sullivan, F, Donnan, P, Thorpe, KE, and Zwarenstein, M. The PRECIS-2 tool: designing trials that are fit for purpose. *BMJ*, 350:h2147, 2015.

8. Ford, I and Norrie, J. Pragmatic trials. *New England Journal of Medicine*, 375(5): 454–463, 2016.

9. FDA. Framework for FDA's Real-World Evidence Program. U.S. Department of Health and Human Services, Food and Drug Administration, Silver Spring, Maryland, USA, 2018.

10. Concato, J, Stein, P, Dal Pan, GJ, Ball, R, and Corrigan-Curay, J. Randomized, observational, interventional, and real-world–What's in a name? *Pharmacoepidemiology and Drug Safety*, 29(11):1514–1517, 2020.

11. ICH. ICH E6 (R3) Guideline for Good Clinical Practice Annex-2. The International Council for Harmonisation of Technical Requirements for Pharmaceuticals for Human Use (ICH), 2023. Accessed: November 20, 2023.

12. Petrini, C, Mannelli, C, Riva, L, Gainotti, S, and Gussoni, G. Decentralised Clinical Trials (DCTs): a few ethical considerations. *Frontiers in Public Health*, 10:1081150, 2022.

13. Warraich, HJ, Marston, HD, and Califf, RM. Addressing the Challenge of Common Chronic Diseases–A View from the FDA. *New England Journal of Medicine*, 390:490–492, 2024.

14. FDA. Digital Health Technologies for Remote Data Acquisition in Clinical Investigations. The US Food and Drug Aministration, Department of Health and Human Services, 2023.

15. FDA. Framework for the Use of Digital Health Technologies in Drug and Biological Product Development. The Food and Drug Administration, 2023. Accessed: March 24, 2024.

16. Goldsack, JC, Coravos, A, Bakker, JP, Bent, B, Dowling, AV, Fitzer-Attas, C, Godfrey, A, Godino, JG, Gujar, N, and Izmailova, E. Verification, analytical validation, and clinical validation (V3): the foundation of determining fit-for-purpose for Biometric Monitoring Technologies (BioMeTs). *npj Digital Medicine*, 3(1):55, 2020.

17. Demanuele, C, Lokker, C, Jhaveri, K, Georgiev, P, Sezgin, E, Geoghegan, C, Zou, KH, Izmailova, E, and McCarthy, M. Considerations for conducting bring your own "Device"(BYOD) Clinical Studies. *Digital Biomarkers*, 6(2):47–60, 2022.

18. Liu, R, Rizzo, S, Whipple, S, Pal, N, Pineda, AL, Lu, M, Arnieri, B, Lu, Y, Capra, W, and Copping, R. Evaluating eligibility criteria of oncology trials using real-world data and AI. *Nature*, 592(7855):629–633, 2021.

19. Thomas, KA and Kidziński, L. Artificial intelligence can improve patients experience in decentralized clinical trials. *Nature Medicine*, 28(12):2462–2463, 2022.

20. Chen, J, Lu, Y, and Kummar, S. Increasing Patient Participation in Oncology Clinical Trials. *Cancer Medicine*, 12(3):2219–2226, 2023.

21. FDA. Using Artificial Intelligence & Machine Learning in the Development of Drug and Biological Products. The US Food and Drug Administration, 2023. Accessed: March 24, 2024.

22. Harmon, DM, Noseworthy, PA, and Yao, X. The Digitization and Decentralization of Clinical Trials. In *Mayo Clinic Proceedings*, volume 98, pages 1568–1578. Elsevier, 2023.

23. Hutson, M. How AI is being used to accelerate clinical trials. *Nature*, 627(8003):S2–S5, 2024.

24. NMPA. Applications of Artificial Intelligence in Regulatory Science. The National Medical Products Administration of China, 2024. Accessed by June 28, 2024.

25. ICH. E9(R1) Statistical Principles for Clinical Trials: Addendum: Estimands and Sensitivity Analysis in Clinical Trials. International Conference on Harmonisation of Technical Requirements for Registration of Pharmaceuticals for Human Use, 2021.

26. Josan, K, Touros, A, Petlura, C, Parameswaran, V, Grewal, U, Senior, M, Viethen, T, Mundl, H, Seninger, C, and Luithle, J. Validation of a pandemic-proof, decentralized cardiovascular trial: scalable design produces rapid recruitment, high engagement and protocol adherence in DeTAP (Decentralized Trial in Afib Patients). *European heart journal*, 42(Supplement 1):ehab724–3177, 2021.

27. Adams, DV, Long, S, and Fleury, ME. Association of remote technology use and other decentralization tools with patient likelihood to enroll in cancer clinical trials. *JAMA Network Open*, 5(7):e2220053, 2022.

28. Sarraju, A, Seninger, C, Parameswaran, V, Petlura, C, Bazouzi, T, Josan, K, Grewal, U, Viethen, T, Mundl, H, and Luithle, J. Pandemic-proof recruitment and engagement in a fully decentralized trial in atrial fibrillation patients (DeTAP). *NPJ Digital Medicine*, 5(1):80, 2022.

29. Daniore, P, Nittas, V, and von Wyl, V. Enrollment and retention of participants in

remote digital health studies: Scoping review and framework proposal. *Journal of Medical Internet Research*, 24(9):e39910, 2022.

30. Orri, M, Lipset, CH, Jacobs, BP, Costello, AJ, and Cummings, SR. Web-based trial to evaluate the efficacy and safety of tolterodine ER 4 mg in participants with overactive bladder: REMOTE trial. *Contemporary Clinical Trials*, 38(2):190–197, 2014.

31. Marquis-Gravel, G, Roe, MT, Turakhia, MP, Boden, W, Temple, R, Sharma, A, Hirshberg, B, Slater, P, Craft, N, and Stockbridge, N. Technology-enabled clinical trials: transforming medical evidence generation. *Circulation*, 140(17):1426–1436, 2019.

32. De Brouwer, W, Patel, CJ, Manrai, AK, Rodriguez-Chavez, IR, and Shah, NR. Empowering clinical research in a decentralized world. *NPJ Digital Medicine*, 4(1):102, 2021.

33. Sehrawat, O, Noseworthy, PA, Siontis, KC, Haddad, TC, Halamka, JD, and Liu, H. Data-Driven and Technology-Enabled Trial Innovations Toward Decentralization of Clinical Trials: Opportunities and Considerations. In *Mayo Clinic Proceedings*, volume 98, pages 1404–1421. Elsevier, 2023.

34. Glazer, JV, MacDonnell, K, Frederick, C, Ingersoll, K, and Ritterband, LM. Liar! Liar! Identifying eligibility fraud by applicants in digital health research. *Internet Interventions*, 25:100401, 2021.

35. Skipper, CP, Pastick, KA, Engen, NW, Bangdiwala, AS, Abassi, M, Lofgren, SM, Williams, DA, Okafor, EC, Pullen, MF, and Nicol, MR. Hydroxychloroquine in nonhospitalized adults with early COVID-19: a randomized trial. *Annals of Internal Medicine*, 173(8):623–631, 2020.

36. EMA. Recommendation paper on decentralised elements in clinical trials, 2022. Accessed: November 20, 2023.

37. Curtis, A and Qu, Y. Impact of using a mixed data collection modality on statistical inferences in decentralized clinical trials. *Therapeutic Innovation & Regulatory Science*, 56(5):744–752, 2022.

38. Steinhubl, SR, Waalen, J, Edwards, AM, Ariniello, LM, Mehta, RR, Ebner, GS, Carter, C, Baca-Motes, K, Felicione, E, and Sarich, T. Effect of a home-based wearable continuous ECG monitoring patch on detection of undiagnosed atrial fibrillation: the mSToPS randomized clinical trial. *JAMA*, 320(2):146–155, 2018.

39. Peigh, G and Passman, RS. "Pill-in-Pocket" anticoagulation for stroke prevention in atrial fibrillation. *Journal of Cardiovascular Electrophysiology*, 34(10):2152–2157, 2023.

40. Paganelli, AI, Mondéjar, AG, da Silva, AC, Silva-Calpa, G, Teixeira, MF, Carvalho, F, Raposo, A, and Endler, M. Real-time data analysis in health monitoring systems: A comprehensive systematic literature review. *Journal of Biomedical Informatics*, 127: 104009, 2022.

41. ICH. Good Clinical Practice E6(R3). The International Council for Harmonisation of Technical Requirements for Pharmaceuticals for Human Use (ICH), 2023. Accessed: November 20, 2023.

42. Rego, S, Henriques, AR, Serra, SS, Costa, T, Rodrigues, AM, and Nunes, F. Methods for the Clinical Validation of Digital Endpoints: Protocol for a Scoping Review Abstract. *JMIR Research Protocols*, 12(1):e47119, 2023.

43. Chen, J, Scharfstein, D, Wang, H, Yu, B, Song, Y, He, W, Scott, J, Lin, X, and Lee, H. Estimands in Real-World Evidence Studies. *Statistics in Biopharmaceutical Research*, 16(2):257–269, 2024.

44. Rockhold, FW, Tenenbaum, JD, Richesson, R, Marsolo, KA, and O'Brien, EC. Design and analytic considerations for using patient-reported health data in pragmatic clinical trials: report from an NIH Collaboratory roundtable. *Journal of the American Medical Informatics Association*, 27(4):634–638, 2020.

45. Izem, R, Zuber, E, Daizadeh, N, Bretz, F, Sverdlov, A, Edrich, P, Branson, J, Degtyarev, E, Sfikas, N, and Hemmings, R. Decentralized clinical trials: scientific considerations through the lens of the estimand framework. *Therapeutic Innovation & Regulatory Science*, 58:495–504, 2024.

46. Adashi, EY and Marks, PW. Achieving Diversity Through Decentralization. *Journal of General Internal Medicine*, pages 1–2, 2024.

47. McCarthy, MW, Lindsell, CJ, Rajasingham, R, Stewart, TG, Boulware, DR, and Naggie, S. Progress toward realizing the promise of decentralized clinical trials. *Journal of Clinical and Translational Science*, 8(1):e19, 2024.

48. Giraudeau, B, Caille, A, Eldridge, SM, Weijer, C, Zwarenstein, M, and Taljaard, M. Heterogeneity in pragmatic randomised trials: sources and management. *BMC Medicine*, 20(1):372, 2022.

49. Leyrat, C, Eldridge, S, Taljaard, M, and Hemming, K. Practical considerations for sample size calculation for cluster randomized trials. *Journal of Epidemiology and Population Health*, 72(1):202198, 2024.

50. NRC. *The prevention and treatment of missing data in clinical trials.* National Research Council, National Academies Press, 2010.

51. EMA. Guideline on Missing Data in Confirmatory Clinical Trials. European Medicines Agency, 2011. Accessed: August 7, 2023.

52. Little, RJ and Rubin, DB. *Statistical analysis with missing data*, volume 793. John Wiley & Sons, 2019.

53. Di, J, Demanuele, C, Kettermann, A, Karahanoglu, FI, Cappelleri, JC, Potter, A, Bury, D, Cedarbaum, JM, and Byrom, B. Considerations to address missing data when deriving clinical trial endpoints from digital health technologies. *Contemporary Clinical Trials*, 113:106661, 2022b.

54. Molenberghs, G and Kenward, M. *Missing data in clinical studies.* John Wiley & Sons, 2007.

55. Dalton, AR, Bottle, A, Soljak, M, Okoro, C, Majeed, A, and Millett, C. The comparison of cardiovascular risk scores using two methods of substituting missing risk factor data in patient medical records. *Informatics in Primary Care*, 19(4):225–232, 2011.

56. Goldstein, BA, Navar, AM, Pencina, MJ, and Ioannidis, JP. Opportunities and challenges in developing risk prediction models with electronic health records data: a systematic review. *Journal of the American Medical Informatics Association*, 24(1):198–208, 2017.

57. Beaulieu-Jones, BK, Lavage, DR, Snyder, JW, Moore, JH, Pendergrass, SA, and Bauer, CR. Characterizing and managing missing structured data in electronic health records: data analysis. *JMIR Medical Informatics*, 6(1):e8960, 2018.

58. Martín-Merino, E, Calderón-Larrañaga, A, Hawley, S, Poblador-Plou, B, Llorente-Garc´ıa, A, Petersen, I, and Prieto-Alhambra, D. The impact of different strategies to handle missing data on both precision and bias in a drug safety study: a multidatabase multinational population-based cohort study. *Clinical Epidemiology*, 10:643–654, 2018.

59. Cesare, N and Were, LP. A multi-step approach to managing missing data in time and patient variant electronic health records. *BMC Research Notes*, 15(1):64, 2022.

60. Pond, G. Statistical issues in the use of dynamic allocation methods for balancing baseline covariates. *British Journal of Cancer*, 104(11):1711–1715, 2011.

61. FDA. *Guidance for Industry and Investigators: Safety Reporting Requirements for INDs and BA/BE Studies*. US Food and Drug Administration, 2012.

62. Di, J, Bai, J, Karahanoglu, F, Camerlingo, N, and Demanuele, C. Deployment and application of multi-modal sensors in clinical trials. *Biopharmaceutical Report*, 29(1): 7–10, 2022a.

63. Zipunnikov, V, Dey, D, Merikangas, K, and Leroux, A. Statistical Challenges of Modeling of Mobile Digital Health Data. *Biopharmaceutical Report*, 29(1):11–14, 2022.

64. Apostolaros, M, Babaian, D, Corneli, A, Forrest, A, Hamre, G, Hewett, J, Podolsky, L, Popat, V, and Randall, P. Legal, regulatory, and practical issues to consider when adopting decentralized clinical trials: recommendations from the clinical trials transformation initiative. *Theuraptic Innovation & Regulatory Science*, 54(4):779–787, 2020.

65. Bierer, BE and White, SA. Ethical Considerations in Decentralized Clinical Trials. *Journal of Bioethical Inquiry*, 20(4):711–718, 2023.

66. Underhill, C, Freeman, J, Dixon, J, Buzza, M, Long, D, Burbury, K, Sabesan, S, McBurnie, J, and Woollett, A. Decentralized Clinical Trials as a New Paradigm of Trial Delivery to Improve Equity of Access. *JAMA Oncology*, 10(4):526–530, 2024.

67. NMPA. Guidance on the Design of Patient-Centric Clinical Trials. The Center for Drug Evaluation, National Medical Products Administration of China, 2023. Accessed: November 20, 2023.

68. NMPA. Guidance on the Implementation of Patient-Centric Clinical Trials. The Center for Drug Evaluation, National Medical Products Administration of China, 2023. Accessed: November 20, 2023.

69. NMPA. Guidance on Benefit-Risk Assessment for Patient-Centric Clinical Trials. The Center for Drug Evaluation, National Medical Products Administration of China, 2023. Accessed: November 20, 2023.

70. PMDA. Q&A on Implementation of Clinical Trials for Pharmaceuticals, Medical Devices, and Regenerative MedicinesProducts during COVID-19. Pharmaceutical and Medical Devices Agency (PMDA), Japan, 2020. Accessed: February 20, 2024.

71. Health Canada. Clinical Trials Modernization: Consultation Paper. Health Canada, Government of Canada, 2021. Accessed: February 20, 2024.

72. SMPA. Decentralised Clinical Trials. Swedish Medical Products Agency (SMPA), Sweden, 2020. Accessed: February 20, 2024.

73. DMA. The Danish Medicines Agency's guidance on the implementation of decentralised elements in clinical trials with medicinal products. Danish Medecines Agency, Denmark, 2021. Accessed: February 20, 2024.

74. Swissmedic. Position Paper on decentralised clinical trials (DCTs) with medicinal products in Switzerland. Swissmedic, Bern, Switzerland, 2022. Accessed: February 20, 2024.